\documentclass[a4paper,12pt]{article}

\usepackage[utf8]{inputenc}
\usepackage{comment}
\usepackage[margin=2.5cm]{geometry}
\usepackage{graphicx}
\usepackage{caption}
\usepackage{subfig}
\usepackage{hyperref}
\usepackage{amsmath}
\usepackage{xcolor}
\usepackage{authblk}

\usepackage{multirow}

\title{Scalar Quasinormal Modes of Rotating Regular Black Holes}

\author{Fech Scen Khoo\thanks{\href{mailto:fkhoo@ucm.es}{fkhoo@ucm.es}}}
\affil{Departamento de F\'isica Te\'orica and IPARCOS, Facultad de Ciencias F\'isicas, Universidad Complutense de Madrid, 28040 Madrid, Spain}
\date{\today}

\begin{document}

\maketitle

\begin{abstract}
Quasinormal modes are characteristic signatures of compact objects.
Here we consider rotating regular black holes, representing rotating generalizations of the Simpson and Visser metric.
We present the spectrum of scalar quasinormal modes and compare it
with the spectrum of Kerr black holes. 
The calculations are done using a spectral decomposition method.
The scalar modes smoothly connect to the Kerr limit.
Interestingly, 
for particularly low scaled Hawking temperatures, 
the dependence of the modes changes as rotation increases. 
A further
investigation shows that
the real part of the fundamental and first excited modes of 
the static and slowly rotating black holes (about $13\%$ of the extremal angular momentum) progresses in an opposite behavior in this low temperature region.
Meanwhile the imaginary part of the modes crosses 
where the excited modes become longer lived than the fundamental modes.
Rapid rotation however suppresses such tendency.

\end{abstract}

\section{Introduction}

So far General Relativity (GR) has not only passed all solar system tests \cite{Will:2018bme}, but also all observations related to strong gravity, including binary pulsars \cite{Freire:2024adf}, the shadows of the supermassive black holes M87$^*$ and SgrA$^*$  \cite{EventHorizonTelescope:2019dse,EventHorizonTelescope:2022wkp}, and gravitational waves from merger events \cite{LIGOScientific:2016aoc,LIGOScientific:2017vwq,Cahillane:2022pqm}.
Still, there are significant reasons to expect that GR will be superseded by a more fundamental theory of gravity.
In particular, the dark sector in cosmology and the incompatibility of GR with the principles of quantum mechanics have led to a surge of interest in alternative theories of gravity as well as approaches to quantum gravity.

The black holes of GR are very special.
They are uniquely determined by their global charges, i.e., they carry no `hair' \cite{Chrusciel:2012jk}.
But they are plagued by singularities in their interior, where the GR equations then break down, a feature expected to be remedied by a more fundamental theory.
This shortcoming of GR has thus led to the long-standing quest for black holes with regular interior, as pursued early on by Bardeen \cite{Bardeen:1968}, Dymnikova \cite{Dymnikova:1992ux}, Ayon-Beato and Garcia \cite{Ayon-Beato:1998hmi}, Bonanno and Reuter \cite{Bonanno:2000ep}, and Hayward \cite{Hayward:2005gi}.
While at first only spherically symmetric regular black holes have been considered, later also rotating regular black holes have been obtained, e.g. \cite{Bambi:2013ufa,Ghosh:2014pba,Jusufi:2019caq,Guerrero:2020azx,Mazza:2021rgq,Shaikh:2021yux,Lima:2021las,Simpson:2021dyo,Franzin:2022wai,Ghosh:2022gka,Frolov:2023jvi} (see also  \cite{Torres:2022twv,Lan:2023cvz,Frolov:2024hhe,Carballo-Rubio:2025fnc} and references therein).

Of utmost importance is, of course, the investigation of possible observable effects associated with such regular black hole solutions.
Here currently two main avenues are formed by the calculations of the black hole shadow to be compared with future observations of the black hole shadow \cite{Ayzenberg:2023hfw} and gravitational waves from inspiral, merger and ringdown of black holes with the third generation of gravitational wave observatories on the ground \cite{ET:2019dnz,Reitze:2019iox} and the LISA space mission \cite{LISA:2017pwj,Colpi:2024xhw}.
For related studies, see e.g. \cite{
Bronnikov:2012ch, Dymnikova:2015yma, 
Anacleto:2021qoe,
Campos:2021sff,
Konoplya:2022hll, Vagnozzi:2022moj, 
Calza:2022ioe, Konoplya:2023aph, 
Pedrotti:2024znu,  Calza:2024fzo, Calza:2024xdh, Spina:2024npx, Konoplya:2025hgp}.
Whereas the inspiral and merger can already be used to put, for instance, stronger bounds on alternative theories of gravity \cite{Julie:2024fwy}, more accurate observations of the ringdown will be needed to discriminate signatures of black holes beyond GR from Kerr black holes \cite{Berti:2025hly}.

Here we address the ringdown of rotating regular black holes to look for specific telltale signs.
In particular, we consider the rotating generalization of the Simpson-Visser (SV) metric \cite{Simpson:2018tsi}, obtained before \cite{Mazza:2021rgq,Shaikh:2021yux,Lima:2021las} with the method of Azreg-A\"inou \cite{Azreg-Ainou:2014pra}.
We obtain the spectrum of the scalar quasinormal modes of these rotating regular black holes by employing our recently developed method of spectral decomposition of the field perturbations in a rotating background \cite{Blazquez-Salcedo:2023hwg,Khoo:2024yeh,Blazquez-Salcedo:2024oek,Khoo:2024agm,Blazquez-Salcedo:2024dur}.
The quasinormal modes of the non-rotating SV black hole have been studied 
before \cite{Jha:2023wzo,Franzin:2023slm,Malik:2024qsz}.

In section 2 we discuss the rotating black hole background metric \cite{Mazza:2021rgq,Shaikh:2021yux,Lima:2021las}.
Section 3 defines the scalar perturbations and derives the perturbation equation and the boundary conditions.
The results are presented in section 4, where the scalar $(l=2)$-led fundamental and first excited modes are discussed and compared to the limiting non-rotating case and the Kerr case of GR.
The paper ends with conclusions in section 5.
The Appendix discusses the static case \cite{Jha:2023wzo,Franzin:2023slm,Malik:2024qsz},
and also provides the quasinormal mode values of several rotating black hole configurations.

\section{Rotating background}

We consider the following stationary and axially symmetric SV spacetime in Boyer-Lindquist-like coordinates in $(t,R,\theta,\varphi)$ \cite{Lima:2021las},

\begin{eqnarray}
  &&  ds^2=-\left(1-\frac{2MR}{\Sigma}\right) dt^2 + \frac{\Sigma}{\Delta} \frac{dR^2}{B_{SV}(R)}
    - \frac{4 M a R \sin^2 \theta}{\Sigma} dt d\varphi 
    + \Sigma d\theta^2 
    \nonumber \\
   && \quad \quad \;\; +
    \Sigma \left(1+\left(\frac{a\sin\theta}{\Sigma}\right)^2 (\Sigma + 2 M R) \right) \sin^2\theta \, d\varphi^2 \, ,
    \label{rot_SV}
\end{eqnarray}
where
  \begin{eqnarray} 
      \Sigma &=& R^2 + a^2 \cos 
      ^2\theta \, , 
      \nonumber \\   
\Delta &=& R^2 - 2 MR + a^2 \, , 
      \nonumber \\ 
B_{SV}(R) &=& \left( 1 - 
\frac{b^2}{R^2}\right) \, .
  \end{eqnarray}
$M$ is the Arnowitt-Deser-Misner (ADM) mass.
The metric carries two more parameters, $a$ and $b$. 
The parameter
$a$ is  
related to the angular momentum, $J=aM$,
while for the additional parameter $b$ in this background metric,
depending on its value, different geometries result.

The metric has a number of limiting configurations.
For $a=0$ the static spherically symmetric SV spacetime \cite{Simpson:2018tsi} is recovered,
\begin{eqnarray}
    ds^2 = -\left(1-\frac{2M}{R} \right) dt^2 +
    \frac{dR^2}{(1-2M/R)B_{SV}(R)} + R^2 (d\theta^2+\sin^2\theta d\varphi^2) \, ,
    \label{R_stat}
\end{eqnarray}
and when $b=0$, the rotating SV metric (\ref{rot_SV}) reduces to the Kerr metric of GR.
The
Killing horizons are given by
$R_{\pm} = M \pm \sqrt{M^2-a^2}$.
For $b>0$, the metric results in different types of geometries depending on certain conditions:
\begin{eqnarray}
\text{Black hole:}&& \quad M+\sqrt{M^2-a^2} > b 
\quad \text{and} \quad  a<M
\label{bh}
\\
\text{Extremal black hole:}&& \quad a > b 
\quad \text{and} \quad  a=M
\label{ebh}
\\
\text{Null wormhole:}&& \quad M+\sqrt{M^2-a^2} = b
\quad \text{and} \quad  a<M 
\label{wh}
\end{eqnarray}
In addition, ordinary wormhole solutions arise for $b>R_{+}$. 
For our study, we are interested in the regular non-extremal black holes with outer
horizon $R_H=R_+$. 
Note that these regular black holes can also be distinguished by their structure inside the outer horizon, depending if $R_->b$ (existence of an inner horizon) or if $R_-\le b$ (no inner horizon). See reference \cite{Mazza:2021rgq} for a full study of the structure of these geometries inside the outer horizon.

\begin{figure}[h!]%
\begin{center} \includegraphics[width=8.0cm,angle=-90]{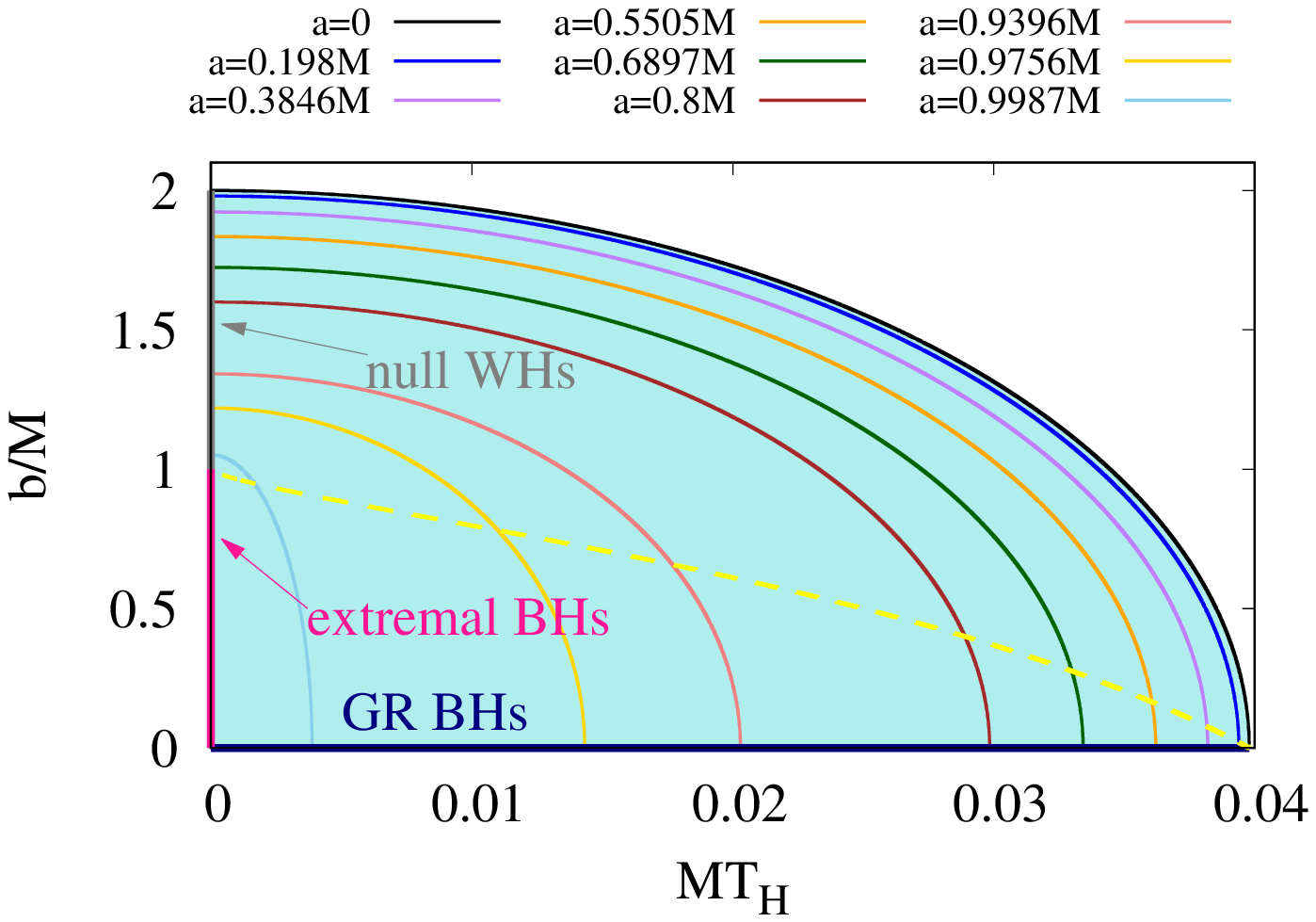}
  \caption{Relation between scaled parameter $b$ and
  scaled Hawking temperature $T_H$ for
  various values of the rotation parameter $a$.
  The limit $b=0$ in dark blue represents the set of black holes (BHs) in GR. 
  In other limits, a set of null wormholes (WHs) is shown in dark gray, the extremal black holes in dark pink, and the null black holes ($R_- = b$, $a<M$) are shown by a yellow dashed curve.  
}
  \label{fig1}
\end{center}
\end{figure}
In figure \ref{fig1}, we show
the relation of the parameter $b$ of the rotating metric and the Hawking temperature $T_H$, both scaled with the black hole mass $M$.
The horizontal line in dark blue represents black holes in GR where extremality is approached towards zero temperature, and the right end of the line is the location of the Schwarzschild black hole.
The set of extremal black holes is represented by the dark pink vertical line,
black holes with a null throat are represented by the yellow dashed curve,
while the dark gray vertical line represents the set of null wormholes. 
These geometries are defined by different relations between the parameters. See conditions (\ref{bh}) - (\ref{wh}) where the parameter $b$ from the SV metric plays its role.

For instance, in the static case $a=0$, a wormhole results
for $b/M=2$, while for 
$b/M<2$ there are black hole geometries (solid black line). 
For $a/M=0.8$, a wormhole is obtained for $b/M=1.6$.
Each curve in the figure corresponds to a family of black holes with a specific value of the rotation parameter $a$, here shown for a set of values of $a$ from the static ($a=0$) up to the almost extremal ($a=0.9987M$) cases. 
These curves are given by
\begin{eqnarray}
    \frac{b}{M} = \frac{2 R_H^2}{R_H^4 - a^4}\sqrt{R_H^4-64R_H^4(\pi MT_H)^2+a^2(a^2-2R_H^2)} \, .  
\end{eqnarray}
The curves become steeper as the black hole rotates faster.
To summarize, for non-extremal black holes, the ranges of the parameters are $0 \le a/M < 1$
and $0 \le b/R_H < 1$. These form the shaded area in figure \ref{fig1}.

\section{Scalar perturbations}

We now consider the scalar field equation,
\begin{eqnarray}
 \mathcal{S} =\frac{1}{\sqrt{-g}} \partial_{\mu} (\sqrt{-g} g^{\mu\nu} \partial_{\nu} \phi)  = 0 \, .
 \label{phi_fe}
\end{eqnarray}
We linearly perturb the scalar field in the rotating SV background ($bg$),
and harmonically decompose the perturbations $\delta \phi$,
\begin{eqnarray}
\phi &=& \phi^{(bg)} + \epsilon \delta\phi(t,R,\theta,\varphi) =  \epsilon e^{i(M_z\varphi-\omega t)} \Phi(R,\theta) \, ,
\label{phi_pert}
\end{eqnarray}
where $M_z$ is the azimuthal number and $\omega$ is the eigenfrequency. 
Plugging equation (\ref{phi_pert}) into
equation (\ref{phi_fe}), we end up with a partial differential equation (PDE) to solve for the scalar perturbations $\Phi(R,\theta)$.

We choose a coordinate compactification
for $R,\theta$ expressed in $x,y$,
\begin{eqnarray}
    x = 1- \frac{R_H}{R} \, , \quad y= \cos \theta \, ,
\end{eqnarray}
where $R_H$ is the location of the horizon.
The scalar perturbation equation in the compactified coordinate is given by,
\begin{eqnarray}
&&  \frac{\partial^2}{\partial x^2} \Phi(x,y)
  =
  \frac{1}{(b(x-1))^2-R_H^2} \times 
  \nonumber \\ &&
  \Big\{
  \frac{1}{(x-1)^4 x^2 (a^2(x-1)+R_H^2)^2}
  \Big(
  \big(R_H^4 + 
  a^2R_H^2(x^3y^2-2x^2y^2-x^3+xy^2+4x^2-5x+2)
   \nonumber \\ &&
  + a^4 (x^4y^2-3x^3y^2+3x^2y^2-x^3-xy^2+3x^2-3x+1) \big)  R_H^4 \omega^2 \Phi(x,y)
  \Big)
   \nonumber \\ &&
  + \frac{4a\omega R_H^4 (a^2+R_H^2)\Phi(x,y)}{x^2(x-1)(a^2(x-1)+R_H^2)^2}
   + \frac{R_H^4(1-y^2)}{x(x-1)^2(a^2(x-1)+R_H^2)} \frac{\partial^2}{\partial y^2} \Phi(x,y)
  \nonumber \\ &&
 + \frac{1}{x(a^2(x-1)+R_H^2)}
 \frac{\partial}{\partial x} \Phi(x,y) \times 
  \nonumber \\ &&
\{b^2a^2(-3x^3+7x^2-5x+1) 
+b^2R_H^2(-2x^2+3x-1)  
+ a^2R_H^2(2x-1)  + R_H^4 \} 
   \nonumber \\ &&
  -\frac{2R_H^4}{x^2(x-1)^2(a^2(x-1)+R_H^2)^2 (y^2-1)} 
  \Big(xy(y^2-1)(a^2(x-1)+R_H^2) \frac{\partial}{\partial y} \Phi(x,y) 
   \nonumber \\ &&
  - 2 \Phi(x,y) (a^2y^2x(x-2)+a^2(y^2+x)+xR_H^2-a^2)  \Big)
  \Big\} \, .
  \label{sc_peq}
\end{eqnarray}

To calculate the quasinormal modes, we impose the following boundary conditions.
At infinity, the perturbations are purely outgoing, given in the following form in $R$ and $\theta$, 
\begin{equation}
    \Phi = \frac{1}{R} e^{i\omega R_*}
    \left(\Phi_1^+(\theta) + \mathcal{O}\left(\frac{1}{R} \right) \right) \, ,
\end{equation}
with
\begin{eqnarray}
\frac{dR_*}{dR} = 1+ \frac{2M}{R} + \mathcal{O}\left(\frac{1}{R^2} \right) \, .
\end{eqnarray}
At the horizon, the perturbations are purely ingoing.
They can be expanded in
\begin{eqnarray}
    \Phi = \frac{1}{R} e^{-i(\omega-M_z\Omega_H) R_*}
    \left(\Phi_1^-(\theta) + \mathcal{O}(R-R_H) \right) \, ,
\end{eqnarray}
with the horizon angular velocity $\Omega_H$.
As noted before (see e.g.
\cite{Blazquez-Salcedo:2024dur})
the relation for the radial function $R_*$ depends on the surface gravity $\kappa_H$ and thus Hawking temperature $T_H$,
\begin{eqnarray}
    \frac{dR_*}{dR} = \frac{g_1}{(R-R_H)} + \mathcal{O}(1) \, ,
\end{eqnarray}
where
\begin{eqnarray}
    g_1 = \frac{1}{4} \kappa_H^{-1} = 
    \frac{1}{8\pi} T_H^{-1} \, .
\end{eqnarray}
On top of that, we also require regularity of the perturbation function on the symmetry axis $\theta=0,\pi$.
All the expansions when substituted in the perturbation equation (\ref{sc_peq}) form a set of boundary conditions.
Next we decompose the perturbation function 
in Chebyshev polynomials of the first kind and Legendre polynomials of the first kind. 
The decomposition transforms the initial PDE to an algebraic equation for the coefficients of the perturbation decomposition.
The resulting eigenvalue problem is then solved using Maple and Matlab with the Multiprecision Computing Toolbox Advanpix \cite{Advanpix}. 
For details of the spectral decomposition method, see
\cite{Blazquez-Salcedo:2023hwg}.

\section{Results}

In this section we present the results of the $(l=2)$-led scalar quasinormal modes for the rotating SV black holes for $M_z=2$.
We always compare our results with the corresponding modes of the Kerr black holes in GR that constitute the limit of $b=0$, 
while we address the static limit of the SV black holes in the Appendix.

In figure \ref{alla_wR}, we show the real part of the frequency of the fundamental mode $\omega_R$
as a function of the Hawking temperature $T_H$, both scaled with the black hole mass $M$, for most of the sets of solutions exhibited in figure \ref{fig1}.
Thus we exhibit the dependence for various values of the rotation parameter $a/M$, from the static case $a=0$ to $a/M=0.9756$.
The gray curve ($b=0$) shows the GR values for the frequencies, obtained using the Teukolsky equation \cite{Berti:2009kk}.
These correspond to the modes of a test scalar field in GR. 
The SV modes smoothly emerge from these GR values, as the background parameter $b$ increases.
In fact the real parts $M \omega_R$ are almost independent of the scaled Hawking temperature, 
however for small $MT_H$ the dependence on the scaled Hawking temperature is prominent.

In general, the scaled real part of the frequency 
is larger for black holes with larger rotation parameter $a/M$, since the GR value is basically retained for most of the $b/R_H$ interval.
Note, that the figure also exhibits thin curves for fixed values of $b/R_H=0.5, 0.9, 0.975$.
When the scaled Hawking temperature becomes very low (less than 0.01), a change occurs.
This change signifies not only the sound deviations of the modes from GR, but also 
interestingly, a difference between black holes with a certain angular momentum.
In particular, there the mode dependence changes from static and 
slowly rotating cases ($a/M=0,0.04997,0.09975$) to the case $a/M=0.1492$.  
For the static and slowly rotating black holes,
the scaled real part of the frequency increases before decreasing again, unlike for values of $a/M=0.1492$ and beyond, where the scaled frequency first decreases before increasing (and decreasing) again.
The more rapidly spinning black holes for $a/M=0.9396, 0.9756$, however, do not show such a dip-and-rise pattern in the dependencies, as their frequencies continue to drop approaching the extremal temperature.

\begin{figure}[h!]%
\begin{center}
  \includegraphics[width=9.0cm,angle=-90]{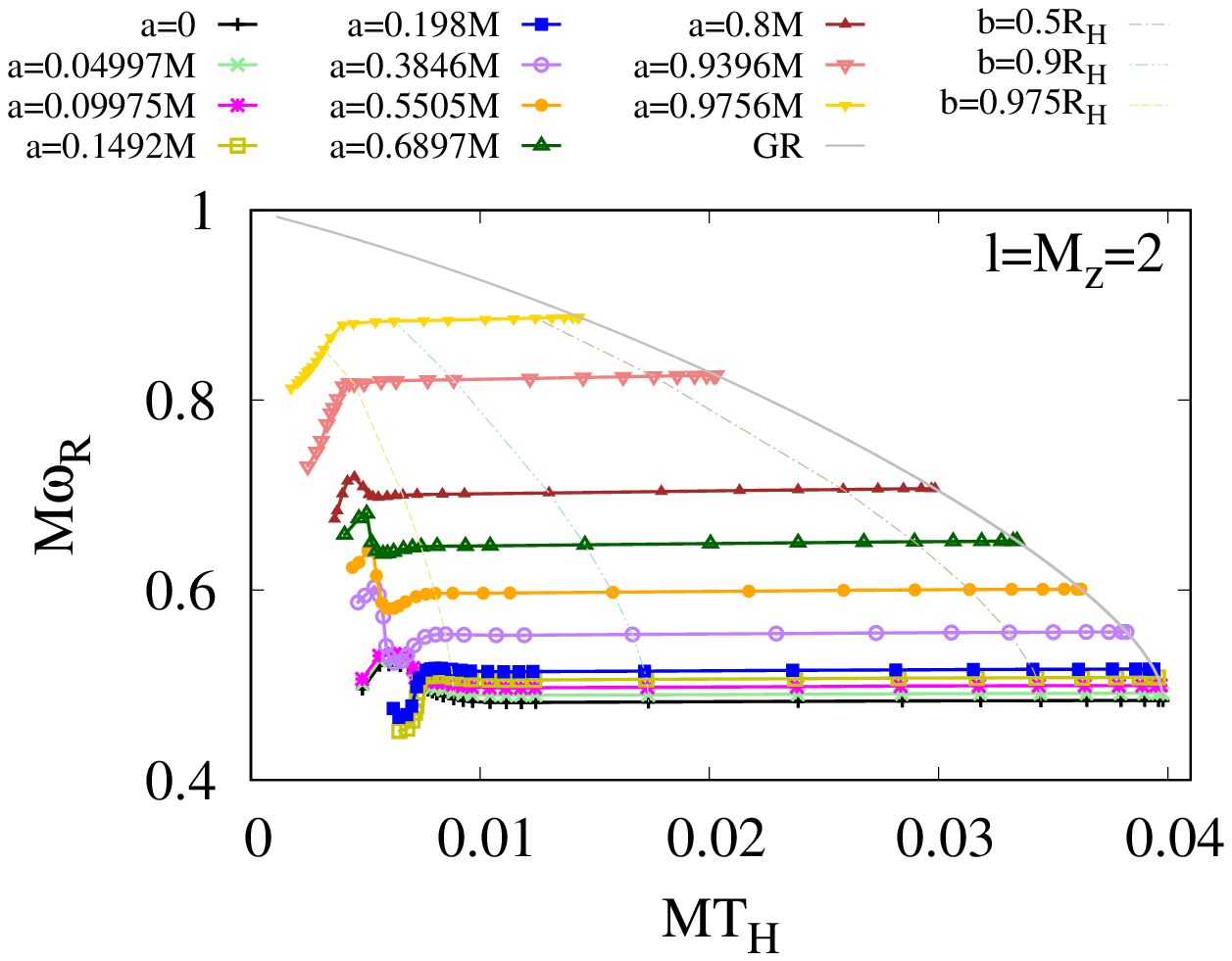}
  \caption{$(l=2)$-led fundamental modes for $M_z=2$: scaled real frequency $M\omega_R$ as a function of the scaled Hawking temperature $MT_H$ for rotation parameter $a/M=0,0.04997,0.09975,0.1492,0.198,0.3846,0.5505,0.6897,0.8,0.9396,0.9756$.
 The GR limit ($b=0$)  given by the Teukolsky equation is shown by a gray curve, and 
 SV black holes with parameter $b/R_H=0.5, 0.9, 0.975$ are shown by thin curves correspondingly. }
  \label{alla_wR}
\end{center}
\end{figure}

\begin{figure}[h!]%
\begin{center}
  \includegraphics[width=8.0cm,angle=-90]{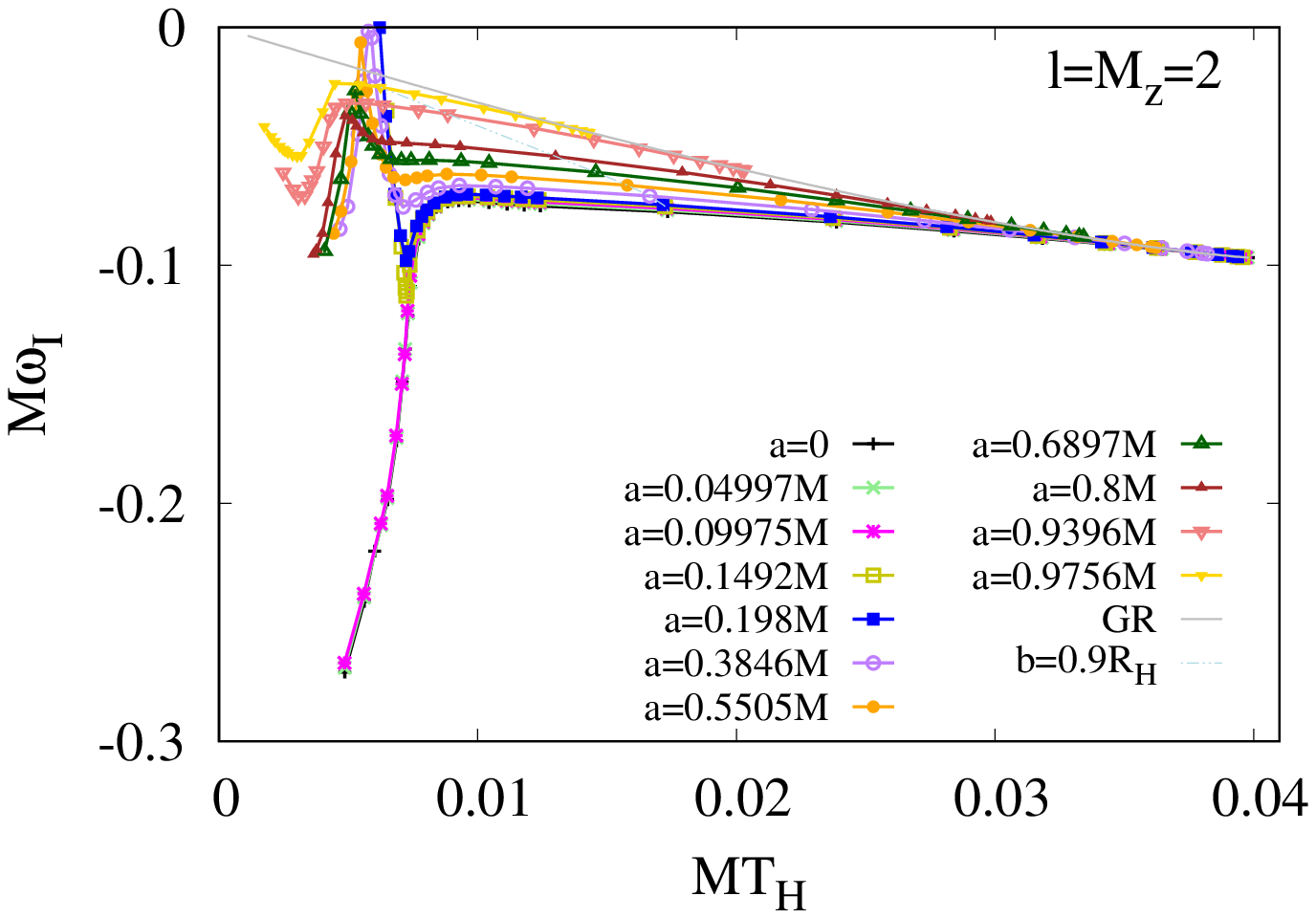}
 \caption{$(l=2)$-led fundamental modes for $M_z=2$: scaled imaginary part of the mode $M\omega_I$ as a function of the scaled Hawking temperature $MT_H$ for rotation parameter $a/M=0,0.04997,0.09975,0.1492,0.198,0.3846,0.5505,0.6897,0.8,0.9396,0.9756$.
 The GR limit ($b=0$) given by the Teukolsky equation is shown by a gray curve, 
 and SV black holes with parameter $b/R_H=0.9$ are shown by a light blue curve. }
  \label{alla_wI}
\end{center}
\end{figure}

\begin{figure}[h!]%
\begin{center}
  \subfloat[]{
   \includegraphics[width=5.5cm,angle=-90]{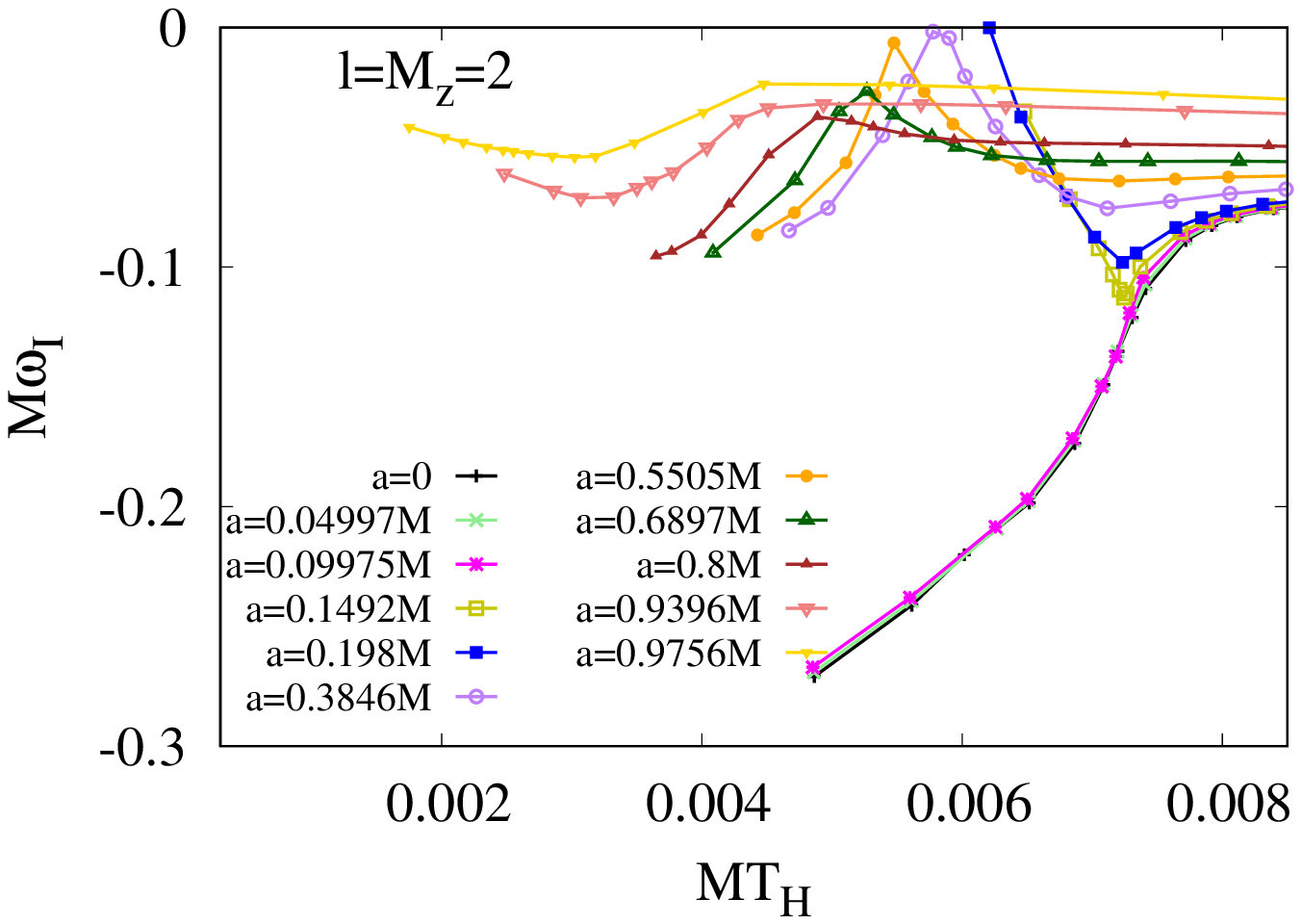}}
   \subfloat[]{
  \includegraphics[width=5.5cm,angle=-90]{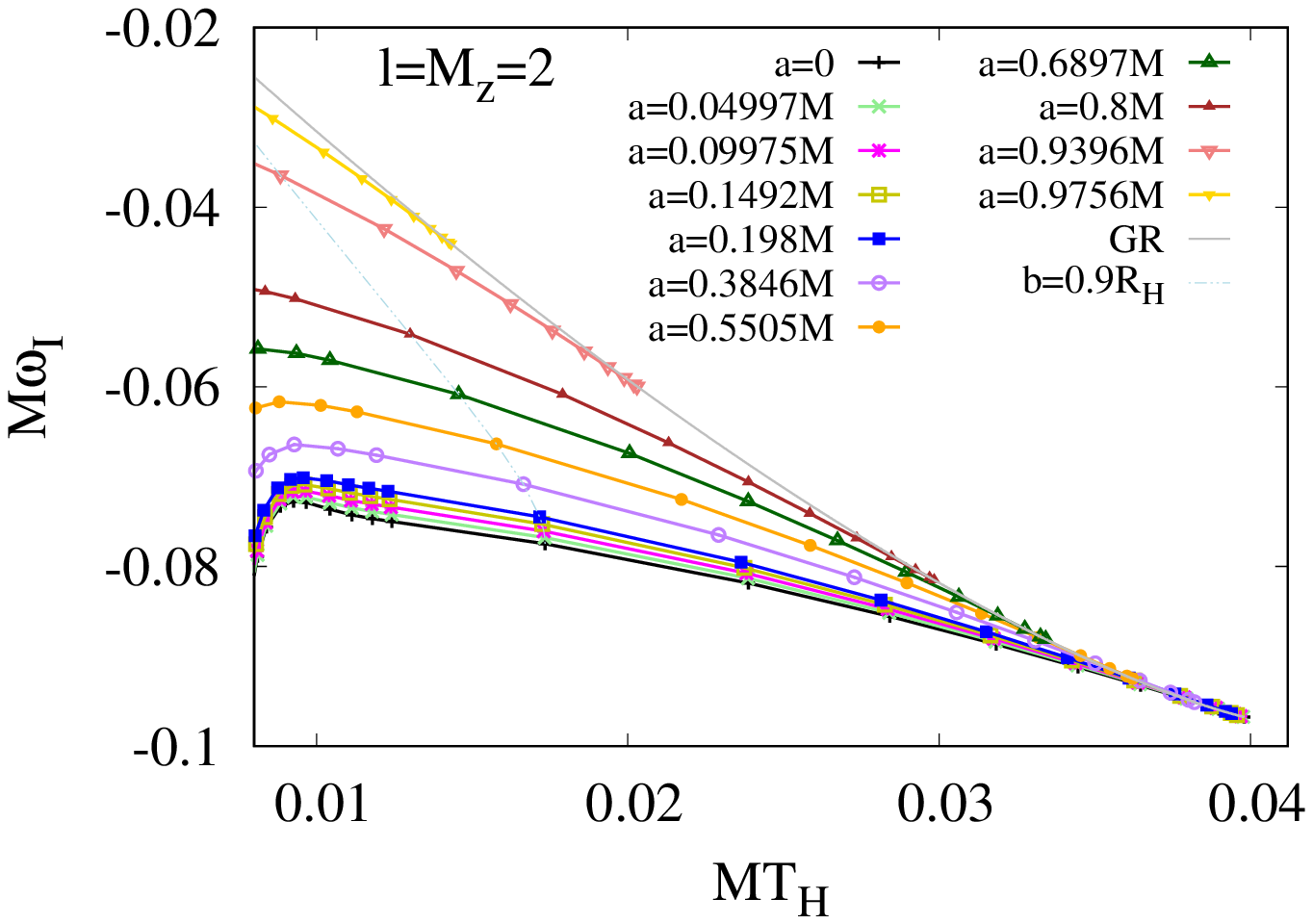}}
  \caption{A close-up of the scaled imaginary part of the scalar modes shown in figure \ref{alla_wI},
  for the (a) lower and (b) higher scaled Hawking temperatures.}
  \label{alla_wI_zoom}
\end{center}
\end{figure}

We show in figure \ref{alla_wI} the scaled imaginary part of the fundamental mode $M\omega_I$ as a function of the scaled Hawking temperature $MT_H$, for the same values of $a/M$.
Recall that $\omega_I$ is inversely proportional to the damping time of the mode.
Zooming in, we exhibit separately
these results for the scaled Hawking temperature between 0.002 to 0.008 (figure \ref{alla_wI_zoom}(a)), and between 0.01 to 0.04 (figure \ref{alla_wI_zoom}(b)).

Whereas the scaled real part remains basically constant, the scaled imaginary part of the modes 
is more sensitive to the Hawking temperature. 
It
decreases almost linearly with increasing scaled Hawking temperature for most of the $b/M$ interval.
In this region of higher temperature, the absolute value of the scaled imaginary part is smaller for black holes that spin faster, as for the Kerr black holes, where the curves emerge.
This almost linear behavior changes again,
like the real part of the modes,
when the lower Hawking temperature, 
that is when the extremal limit
is approached.
As already noted for the real part of the modes,
when $a/M=0,0.04997,0.09975$, the modes behave quite differently as compared to the modes of higher spinning SV black holes.
The modes for $a/M=0,0.04997,0.09975$ reach a maximal value and then decrease rapidly, while the modes for the parameters $a/M=0.1492, 0.198$  decrease only a little after reaching a (local) maximum and then rise steeply.
Intriguingly, these modes seem to become more unstable as $b/R_H$ grows, 
where for $a/M=0.198$ the value of the scaled imaginary part reaches around $-8\times 10^{-5}$ for $b/R_H=0.9875$.
Due to the lack of precision, we stop tracking the modes beyond this point for this case. It will be interesting to clarify in the future if these modes truly become unstable and whether they would then regain stability as $b/R_H$ increases further.
For more rapid rotation in the range of $a/M=0.3846$ - $0.8$, the modes show a significant rise and fall, and for $a/M \ge 0.9396$, the modes decrease before rising again with decreasing temperatures. 

We have explored almost 
the entire range of the parameters that forms the domain of existence of the rapidly rotating 
non-extremal regular black holes.
Compared to the real part of the scalar modes, the imaginary part of the modes, that is the damping time of the modes, is generally more sensitive to the Hawking temperature. 
Notably, as we approach the extremal limit,
where we reach the scaled temperature as low as around 0.002 for the fastest spinning black holes considered,
the fundamental modes of the regular black holes exhibit significant deviations from Kerr black holes of GR, 
and furthermore, interesting features differentiating the regular black holes of slower or faster spin arise.
Perturbations of the extremal regular black holes will deserve 
a separate detailed study, and 
especially
when compared with the extremal non-regular black holes.

\begin{figure}[h!]%
\begin{center}
 \subfloat[]{
  \includegraphics[width=20cm,angle=-90]{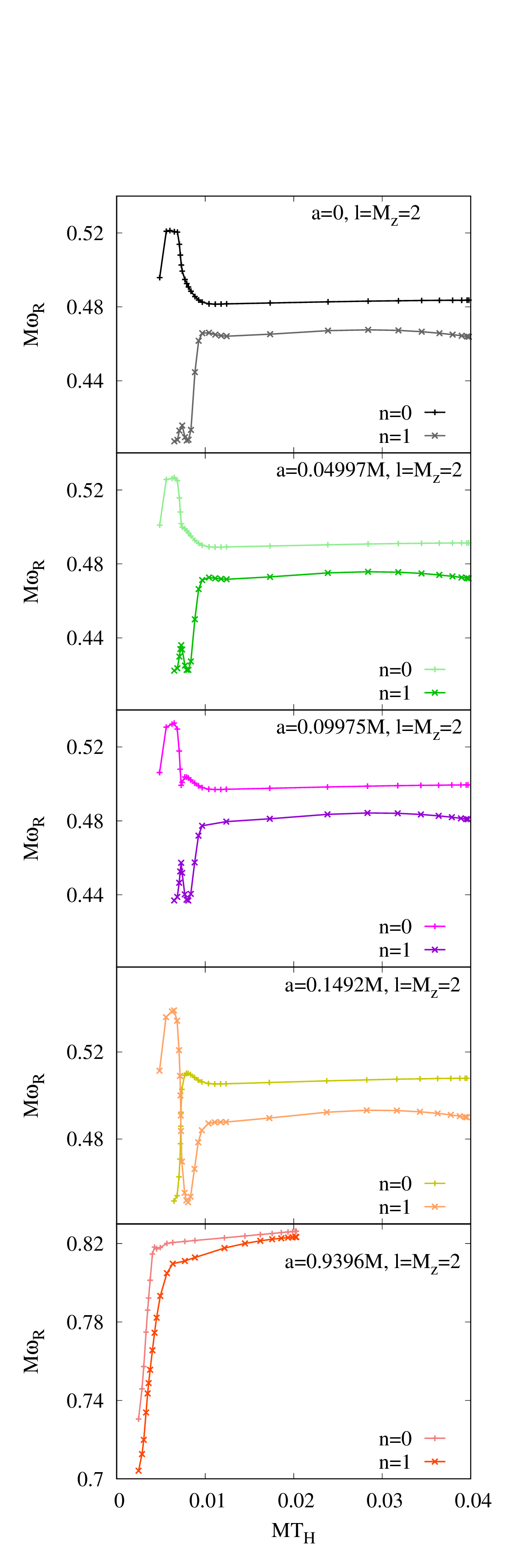}
  }
   \subfloat[]{
  \includegraphics[width=20cm,angle=-90]{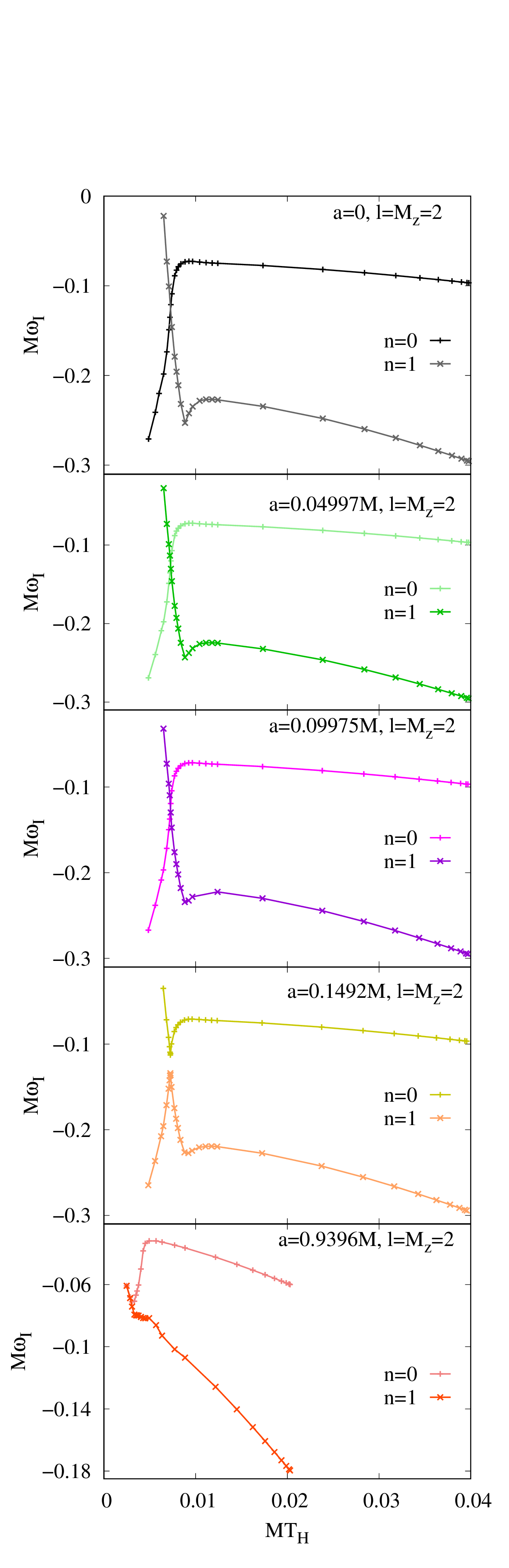}
  }
  \caption{
  $(l=2)$-led fundamental ($n=0$) and first excited ($n=1$) modes for $M_z=2$: (a) scaled real frequency $M\omega_R$ and (b) scaled imaginary part $M\omega_I$ as a function of the scaled Hawking temperature $MT_H$ for the static $(a=0)$ and rotating SV black holes with $a/M=0.04997, 0.09975, 0.1492, 0.9396$ (top to bottom).
  }
  \label{all_a_n0n1}
\end{center}
\end{figure}

The distinctly different behavior of the modes with respect to the scaled Hawking temperature observed when $a/M=0, 0.04997, 0.09975$, as compared to larger $a/M$ values, can be understood through the following results where we compare their fundamental modes ($n=0$) with their first excited modes ($n=1$).
As seen in figure \ref{all_a_n0n1} in the top three rows for $a/M=0, 0.04997, 0.09975$ successively, 
the scaled real parts $M \omega_R$ of the fundamental
and excited modes repel each other, while the scaled imaginary parts $M \omega_I$ of the modes cross.
Consequently, the excited modes become longer lived than the fundamental modes at lower temperatures (i.e., larger $b$).
This \emph{repulsive} behavior in the real parts of the modes diminishes as the rotation increases.
In the fourth row of the figure are the
modes of the black holes with $a/M=0.1492$.
Here, in an opposite fashion, the scaled real parts of the modes cross while the imaginary parts repel.
This means that the fundamental modes are overall longer lived than the excited modes.
The critical change of such behavior depending on the rotation parameter $a/M$ occurs, 
for instance, around $b/R_H=0.9835$ for $a/M=0.09975$.

Next, in the last row of figure \ref{all_a_n0n1} the fundamental
and first excited modes of more rapidly rotating black holes for $a/M=0.9396$ are shown.
Here the scaled real parts of the modes do not cross; both modes increase sharply for low 
temperatures and less, as the temperature rises further. 
The scaled imaginary parts of the modes
cross at very low Hawking temperatures,
but the fundamental and first excited modes 
retain competitive damping times.
In general, for all the cases shown here, the first excited modes behave similarly as the fundamental modes at high scaled temperatures.

\begin{figure}[p!]%
\begin{center}
  \includegraphics[width=5.5cm,angle=-90]{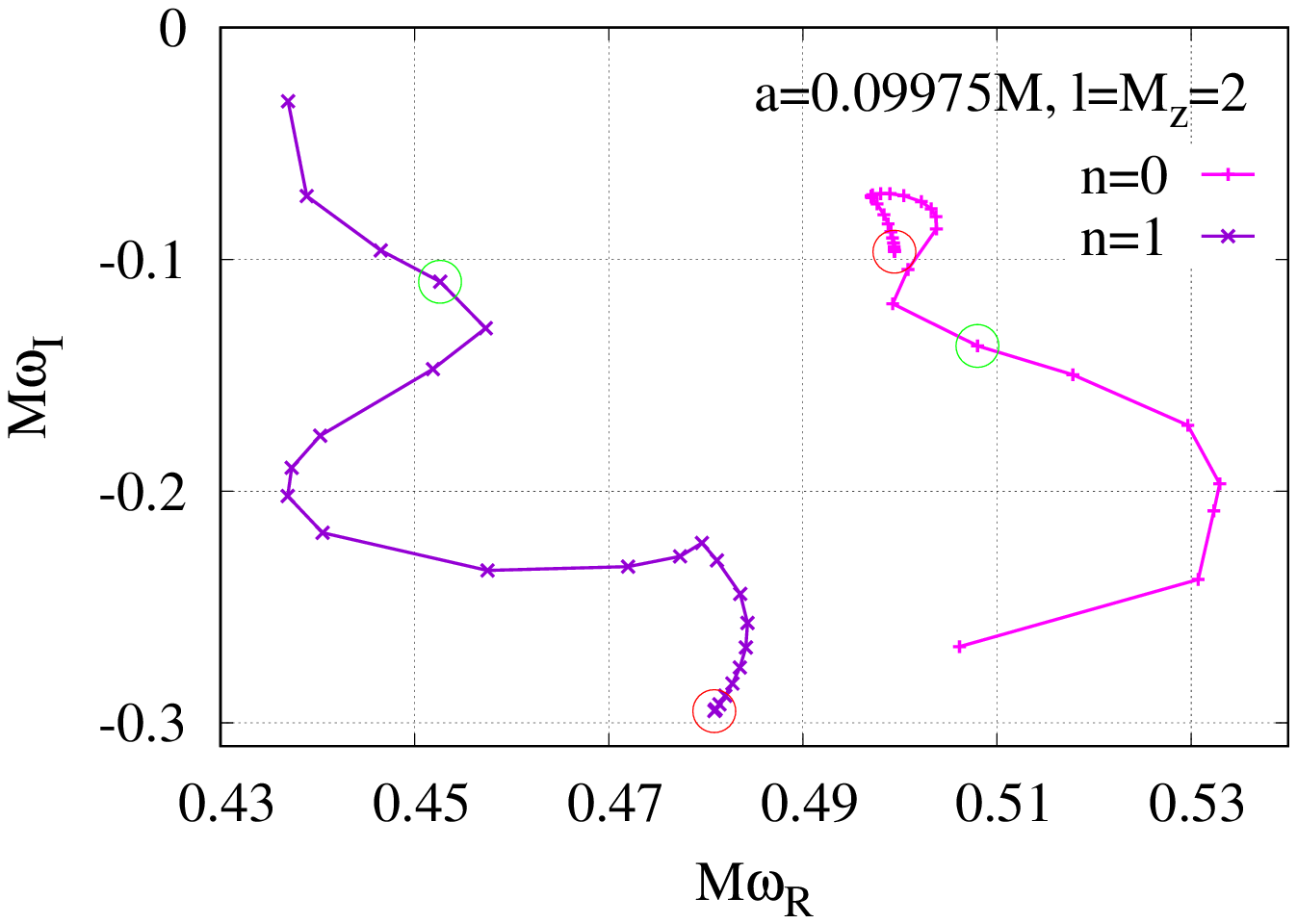}
   \includegraphics[width=5.5cm,angle=-90]{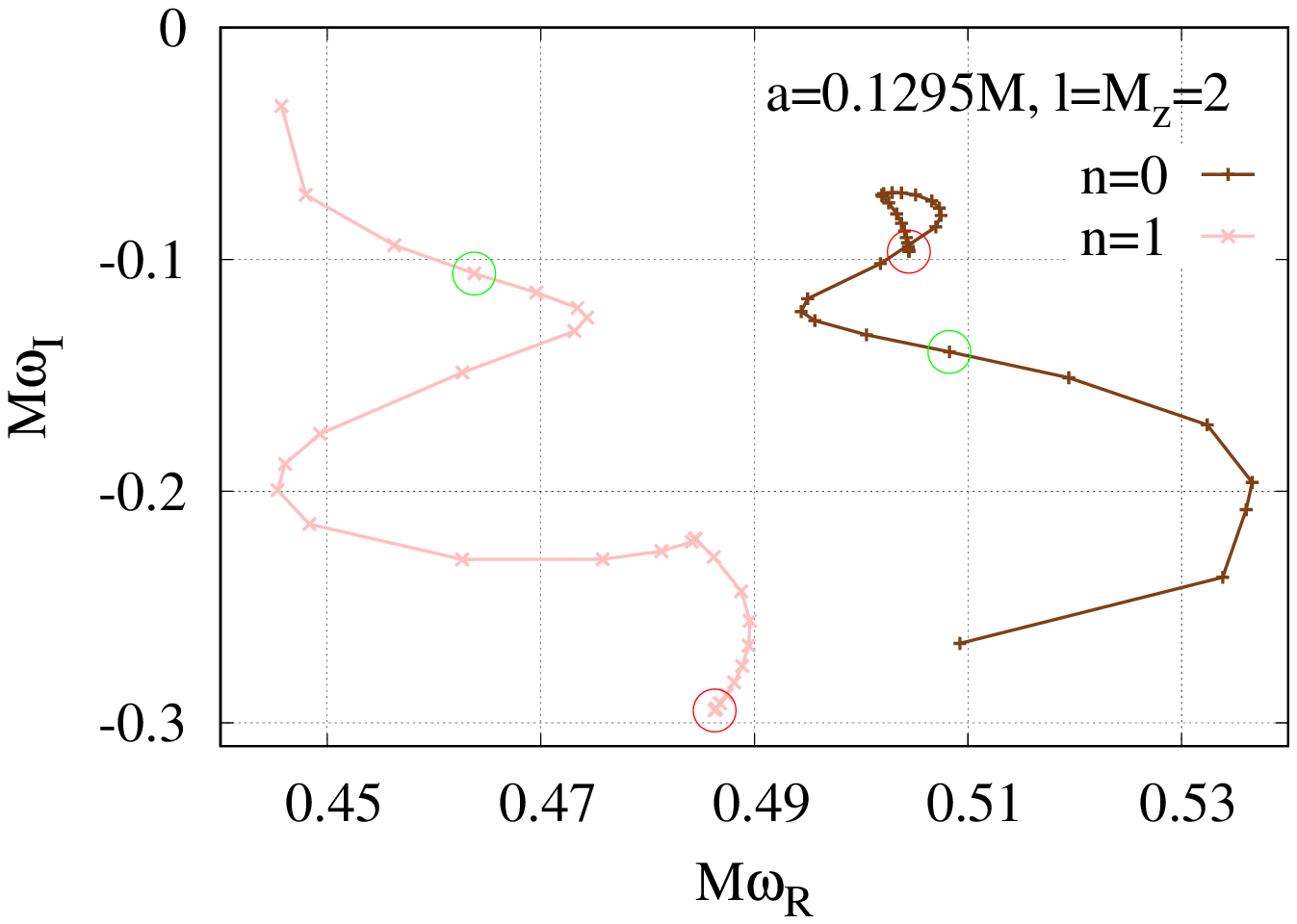}
\includegraphics[width=5.5cm,angle=-90]{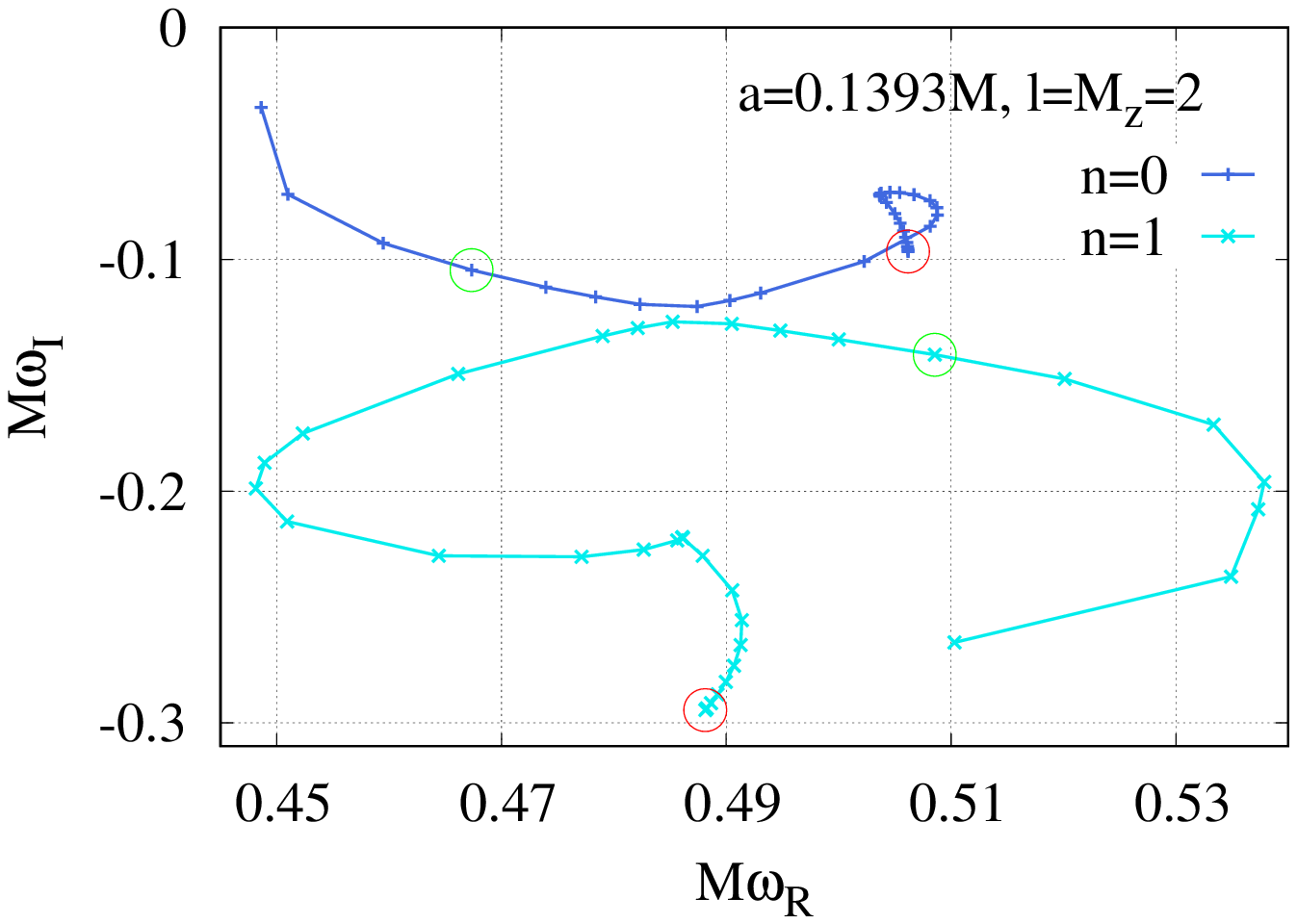}
  \includegraphics[width=5.5cm,angle=-90]{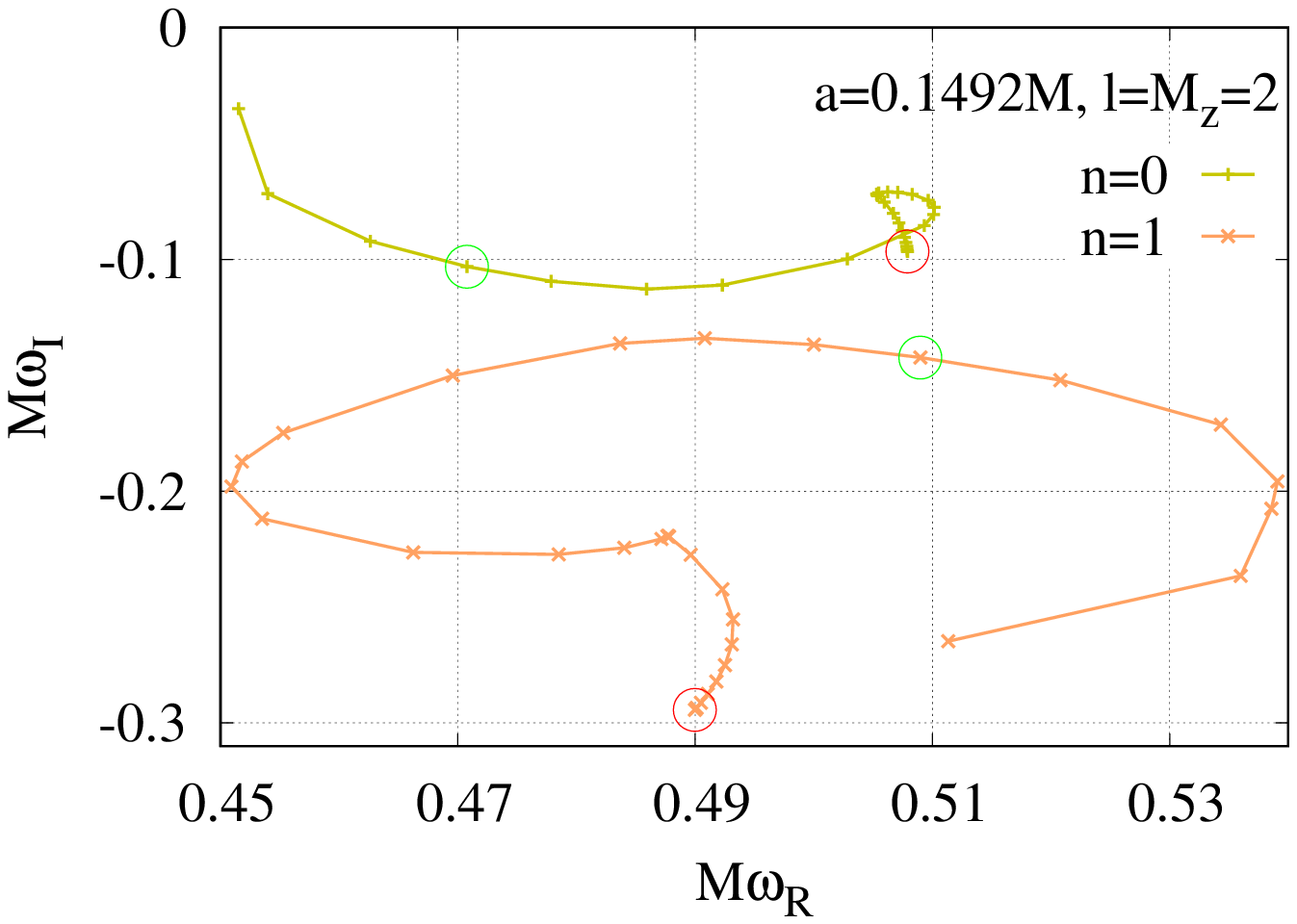}
  \caption{
  $(l=2)$-led fundamental ($n=0$) and first excited ($n=1$) modes for $M_z=2$: scaled imaginary part $M\omega_I$ as a function of the scaled real frequency $M\omega_R$ for the rotating SV black holes with $a/M=0.09975$ (top left), $a/M=0.1295$ (top right), $a/M=0.1393$ (bottom left), and $a/M=0.1492$ (bottom right). Red circles mark the GR Kerr values ($b=0$). Green circles are modes for the parameter $b/R_H=0.9835$.
  }
  \label{a01-a015_n0n1_wRwI}
\end{center}
\end{figure}

We conclude this section by restating the peculiar behavior of the modes
depicted in figure \ref{all_a_n0n1}, 
focusing on
the case of rotation parameter $a/M=0.09975$ 
up to
$a/M=0.1492$.
We show the scaled modes in a complex plane in figure \ref{a01-a015_n0n1_wRwI}. 
GR Kerr values of the modes are indicated by the red circles. 
The points on each curve correspond to a specific value of $b/R_H$.
Figure \ref{a01-a015_n0n1_wRwI}(top left) shows that the continuous fundamental modes ($n=0$) and first excited modes ($n=1$)
exhibit a clear division in $M\omega_R$, around $M\omega_R=0.49$. 
That is, the real parts of the modes do not cross. 
The fundamental modes then continue down to have a shorter damping time, while the opposite happens with the excited modes.  
This occurs around $b/R_H=0.9835$ (green circles).
As rotation increases, as seen in figure \ref{a01-a015_n0n1_wRwI}(top right) the gap close to $M\omega_I=-0.1$ is closing in, and the difference in the values of $M\omega_R$ decreases.
Followed by the bottom left figure, it shows that
the fundamental and excited modes of faster rotating black holes ($a/M=0.1393$) divide in $M\omega_I$, i.e., the imaginary parts do not cross. 
Therefore the fundamental modes remain the longer lived modes. 
This property of the fundamental modes has long been known as the typical behavior of the modes.
However, here for the scalar quasinormal modes, this property breaks down for larger $b/R_H$ and lower $a/M$. 
It is compelling to observe that this relevant property for the studies of quasinormal modes is exhibited for the slower rotating black holes, at extremely low scaled Hawking temperatures, or equivalently, for an extremely large background parameter $b$.
Finally, in figure \ref{a01-a015_n0n1_wRwI}(bottom right) 
for a larger $a/M$,
the division between the modes in $M\omega_I$ grows.

\section{Conclusions}

Previously we have developed the spectral decomposition method to calculate the quasinormal modes of rotating objects and applied it to alternative gravity theories which contain a scalar field and higher curvature terms, considering the impact of the full strength of the theory parameter (the coupling constant) on the quasinormal mode spectrum \cite{Khoo:2024yeh, Blazquez-Salcedo:2024oek, Khoo:2024agm, Blazquez-Salcedo:2024dur}. 
In these calculations the global background solutions  
were obtained by solving the field equations of the theory, and 
the quasinormal modes for all the fields (i.e., the metric and scalar fields)
were calculated by perturbing 
these background solutions. 

Here we have taken another approach. 
Instead of specifying a gravity theory in the first place, for which the background metric is an exact solution,
we have started directly from a well-known rotating metric endowed with a parameter $b$ which reduces to the metric of the GR Kerr black holes when $b$ is zero.
We have then considered a test scalar field on such a parametrized rotating background known as rotating SV black hole, an attractive rotating regular black hole, and calculated the scalar-led quasinormal modes.

We have presented the results for $(l=2)$-led modes for $M_z=2$, with an emphasis on the fundamental modes for various rotation parameters $a/M$, and made a comparison of the fundamental modes with the first excited modes, particularly for
the static, and for several slowly rotating and 
rapidly rotating black holes. 
We have explored almost the entire domain of the parameters $a$ and $b$.
The scalar modes with different rotation parameters $a/M$ smoothly connect to the corresponding modes in GR as the scaled Hawking temperature increases (or, equivalently, as the background parameter $b$ decreases to zero). 

We have found that in most of the parameter range of $b/R_H$ the real parts of the modes are almost constant, whereas the imaginary parts show a slight linear dependence on the scaled Hawking temperature.
The modes exhibit the most distinctive difference from this tendency
when the scaled Hawking temperature is extremely low.
To gain more understanding of this behavior
we have also examined the first excited modes.
Here we have seen that when rotation is very slow
($a/M < 0.1393$) and also when the black holes are static, there arises a \emph{repulsion} between the real parts of the fundamental mode and the first excited mode, while the imaginary parts of the modes cross.
This effectively pushes the first excited modes to be longer-lived than the fundamental modes in these cases. 
Then, from around $a/M=0.1393$ up to around $a/M=0.8$, the imaginary parts of the fundamental and first excited modes remain apart, while
the fundamental modes are always the longest-lived modes. 
For even larger $a/M$, such as $a/M=0.9396$, the real parts of both the fundamental and first excited modes show similar behavior with respect to the scaled Hawking temperature, 
while their imaginary parts almost coincide at the lowest temperatures. 

Our future goal here is to study further rotating regular black holes to see, whether we can find certain sets of observable features that could be attributed to them and that might betray their presence, given sufficiently accurate observations.
Therefore we would like to focus on analyzing the quasinormal mode spectra of rotating regular black holes, derived from specific gravity theories.
In that case, the full spectrum will be obtained including, in particular, the full set of gravitational modes.

\section*{Acknowledgement}

FSK gratefully acknowledges support from
``Atracci\'on de Talento Investigador Cesar Nombela'' of the Comunidad de Madrid, grant no. 2024-T1/COM-31385,
DFG project Ku612/18-1, 
FCT project PTDC/FIS-AST/3041/2020, and
MICINN project PID2021-125617NB-I00 ``QuasiMode''.
FSK is grateful to Jose Luis Bl\'azquez-Salcedo and Jutta Kunz for discussions and comments on the manuscript.

\section*{Appendix}

\subsection*{Static SV black holes}

In this section we present the master equation for the scalar perturbations for static SV black holes ($a=0$), 
\begin{eqnarray}
    \frac{d^2}{dR_*^2} \mathcal{Z} + (\omega^2 - V)  \mathcal{Z} = 0 \, ,
\end{eqnarray}
where $\Phi = \frac{1}{R}\mathcal{Z}(R)Y_{l,M_z}(\theta,\varphi)$ with spherical harmonics $Y_{l,M_z}$,
for two different coordinate systems.
In one coordinate, the expression of the static metric is given in equation (\ref{R_stat}).
The corresponding tortoise coordinate $R_*$ is defined as, 
\begin{eqnarray}
     \frac{dR_*}{dR} = \frac{R^2 }{(R_H-R)\sqrt{R^2-b^2}}\, . 
\end{eqnarray}
In this coordinate, the potential is then given by
\begin{eqnarray}
    V(R) = \frac{1}{R^6} \left((R-R_H)( R^2 (l(l+1) R +  R_H) + b^2 (R - 2R_H)) \right) \, .
\end{eqnarray}

Another way of expressing the static metric in $(t,r,\theta,\varphi)$ is,
\begin{eqnarray}
   ds^2 = - f(r) dt^2 + \frac{dr^2}{f(r)} + (r^2 + b^2) (d\theta^2+\sin^2\theta d\varphi^2) \, , \qquad f(r) = 1 - \frac{2M}{\sqrt{r^2+b^2}}  \, .
\end{eqnarray}
By defining the tortoise coordinate $r_*$ as,
\begin{eqnarray}
    \frac{dr_*}{dr} = \frac{1}{f(r)} \, ,
\end{eqnarray}
the potential term in the master equation for $\mathcal{Z}$ 
is given by
\begin{eqnarray}
    V(r) = \frac{ l(l+1) f(r)}{b^2+r^2} + \frac{f(r) \left( r(b^2+r^2) \frac{d}{dr} f(r) + b^2f(r) \right)}{(b^2+r^2)^2} \, . 
\end{eqnarray}
The relation between the radial coordinates is dictated by $R^2 = r^2 + b^2$, therefore the two expressions of the potential are simply related by a coordinate transformation. 
Here the potential term agrees with the derivation in \cite{Franzin:2023slm} which corrects the potential term shown in \cite{Jha:2023wzo}. Our results agree with \cite{Franzin:2023slm, Malik:2024qsz} up to the point when the mode dependence changes drastically for extremely large $b/R_H$. The novel feature we obtain here is that the first excited modes of the static black holes carry a longer damping time compared to the corresponding fundamental modes for these values of $b/R_H$.

Numerical values of the fundamental and first excited quasinormal modes of the static black holes for $l=M_z=2$ are given in table \ref{tab:a0_n0n1}.
\begin{table}
    \centering
    \begin{tabular}{|c||c|c||c|c||}
\hline
  \multirow{2}{*}{}  &
   \multicolumn{2}{|c||}{$n=0$} &
  \multicolumn{2}{|c||}{$n=1$} \\ 
  $b/R_H$   & $M\omega_R$ & $M\omega_I$  & $M\omega_R$ & $M\omega_I$ \\
 \hline
0	&	0.48364	&	-0.09676	&  0.46385      &	-0.29560 \\
0.2	&	0.48363	&	-0.09589	&  0.46438      &	-0.29284 \\
0.4	&	0.48356	&	-0.09324	&  0.46576      &	-0.28437 \\
0.5	&	0.48348	&	-0.09120	&  0.46658      &	-0.27786 \\
0.6	&	0.48334	&	-0.08865	&  0.46728      &	-0.26970 \\
0.7	&	0.48311	&	-0.08555	&  0.46760      &	-0.25980 \\
0.8	&	0.48273	&	-0.08184	&  0.46713      &	-0.24806 \\
0.9	&	0.48211	&	-0.07748	&  0.46525      &	-0.23454 \\
0.95	&	0.48163	&	-0.07500	&  0.46415      &	-0.22726 \\
0.975 &	0.48560	&	-0.07330		&  0.44469      &  -0.25288 \\
0.9865 & 0.52069		&	-0.19842		&   0.40728     & -0.02213   \\
0.9925	&	0.49587	&	-0.27090	&        &	\\
 \hline

    \end{tabular}
    \caption{$(l=2)$-led fundamental ($n=0$) and first excited ($n=1$) modes with $M_z=2$ for different scaled background parameter $b/R_H$ for the static SV black holes.}
    \label{tab:a0_n0n1}
\end{table}

\subsection*{Numerical values of the scalar quasinormal modes of rotating SV black holes}

Here we present the numerical values of the quasinormal modes of the rotating SV black holes for $l=M_z=2$, for several values of the rotation parameter $a/M$, in tables \ref{tab:a01_n0n1} - \ref{tab:a16_n0}.

\begin{table}
    \centering
    \begin{tabular}{|c||c|c||c|c||}
\hline
  \multirow{2}{*}{}  &
   \multicolumn{2}{|c||}{$n=0$} &
  \multicolumn{2}{|c||}{$n=1$} \\ 
  $b/R_H$   & $M\omega_R$ & $M\omega_I$   & $M\omega_R$ & $M\omega_I$ \\
 \hline
0	&	0.49944	&	-0.09667	&  0.48088    &	-0.29489 \\
0.2	&	0.49942	&	-0.09574	&  0.48139    &	-0.29197	\\
0.4	&	0.49932	&	-0.09292	&  0.48273    &	-0.28301	\\
0.5	&	0.49922	&	-0.09076	&  0.48350    &	-0.27611	\\
0.8	&	0.49835	&	-0.08076	&  0.48355    &	-0.24437	\\
0.95	&	0.49710	&	-0.07338	&  0.47959    &	-0.22236	\\
0.975 &	0.50039	&	-0.07251			&  0.45751    & -0.23424 \\
0.9865 &	0.53294	&	-0.19683			& 0.43696     & -0.03178 \\
0.9925	&	0.50613	&	-0.26708		&      &	\\
 \hline

    \end{tabular}
    \caption{$(l=2)$-led fundamental ($n=0$) and first excited ($n=1$) modes with $M_z=2$ for different scaled background parameter $b/R_H$ for $a/M=0.09975$.}
    \label{tab:a01_n0n1}
\end{table}

\begin{table}
    \centering
    \begin{tabular}{|c||c|c||c|c||}
\hline
  \multirow{2}{*}{}  &
   \multicolumn{2}{|c||}{$n=0$} &
  \multicolumn{2}{|c||}{$n=1$} \\ 
  $b/R_H$   & $M\omega_R$ & $M\omega_I$   & $M\omega_R$ & $M\omega_I$ \\
 \hline
0	&	0.50791	&	-0.09655	&  0.48999    &	-0.29433 \\
0.2	&	0.50788	&	-0.09561	&  0.49049    &	-0.29133	\\
0.4	&	0.50777	&	-0.09271	&  0.49179    &	-0.28214	\\
0.5	&	0.50766	&	-0.09047	&  0.49253    &	-0.27505	\\
0.8	&	0.50671	&	-0.08016	&  0.49231    &	-0.24236	\\
0.95	&	0.50540	&	-0.07252	&  0.48780    &	-0.21972	\\
0.975 &	0.50830	&	-0.07194			&  0.46625    & -0.22638 \\
0.9865 &	0.45152	&	-0.03497			& 0.53908     & -0.19581 \\
0.9925	&		&			&  0.51134    & -0.26476	\\
 \hline

    \end{tabular}
    \caption{$(l=2)$-led fundamental ($n=0$) and first excited ($n=1$) modes with $M_z=2$ for different scaled background parameter $b/R_H$ for $a/M=0.1492$.}
    \label{tab:a015_n0n1}
\end{table}

\begin{table}
    \centering
    \begin{tabular}{|c||c|c||}
\hline
  $b/R_H$   & $M\omega_R$ & $M\omega_I$  \\
 \hline
0	&	0.55594	&	-0.09510		\\
0.2	&	0.55589	&	-0.09404		\\
0.4	&	0.55571	&	-0.09078		\\
0.5	&	0.55553	&	-0.08827		\\
0.8	&	0.55423	&	-0.07649		\\
0.95	&	0.55249	&	-0.06760		\\
0.975 &	0.55341	&	-0.06752		\\
0.9925	&	0.58708	&	-0.08475		\\
 \hline

    \end{tabular}
    \caption{$(l=2)$-led fundamental modes with $M_z=2$ for different scaled background parameter $b/R_H$ for $a/M=0.3846$.}
    \label{tab:a04_n0}
\end{table}

\begin{table}
    \centering
    \begin{tabular}{|c||c|c||}
\hline
  $b/R_H$   & $M\omega_R$ & $M\omega_I$  \\
 \hline
0	&	0.70682	&	-0.08152		\\
0.2	&	0.70669	&	-0.08038		\\
0.4	&	0.70625	&	-0.07684		\\
0.5	&	0.70589	&	-0.07409		\\
0.8	&	0.70380	&	-0.06084		\\
0.95	&	0.70123	&	-0.05019		\\
0.975 &	0.69985	&	-0.04823		\\
0.9925	&	0.67454	&	-0.09547		\\
 \hline

    \end{tabular}
    \caption{$(l=2)$-led fundamental modes with $M_z=2$ for different scaled background parameter $b/R_H$ for $a/M=0.8$.}
    \label{tab:a1_n0}
\end{table}

\begin{table}
    \centering
    \begin{tabular}{|c||c|c||c|c||}
\hline
  \multirow{2}{*}{}  &
   \multicolumn{2}{|c||}{$n=0$} &
  \multicolumn{2}{|c||}{$n=1$} \\ 
  $b/R_H$   & $M\omega_R$ & $M\omega_I$   & $M\omega_R$ & $M\omega_I$ \\
 \hline
0	&	0.82629	&	-0.05993	&  0.82334    &	-0.17959 \\
0.2	&	0.82611	&	-0.05898	&  0.82321    &	-0.17671	\\
0.4	&	0.82557	&	-0.05603	&  0.82270    &	-0.16777	\\
0.5	&	0.82515	&	-0.05372	&  0.82219    &	-0.16075	\\
0.8	&	0.82294	&	-0.04239	&  0.81771    &	-0.12580	\\
0.95	&	0.82055	&	-0.03268	&  0.80972    &	-0.09288	\\
0.975 &	0.81733	&	-0.03354			&  0.78219    & -0.08169 \\
0.9925	&	0.73046	&	-0.06083		&  0.70416    & -0.06079	\\
 \hline

    \end{tabular}
    \caption{$(l=2)$-led fundamental ($n=0$) and first excited ($n=1$) modes with $M_z=2$ for different scaled background parameter $b/R_H$ for $a/M=0.9396$.}
    \label{tab:a14_n0n1}
\end{table}

\begin{table}
    \centering
    \begin{tabular}{|c||c|c||}
\hline
  $b/R_H$   & $M\omega_R$ & $M\omega_I$  \\
 \hline
0	&	0.88751	&	-0.04405		\\
0.2	&	0.88734	&	-0.04330		\\
0.4	&	0.88681	&	-0.04097		\\
0.5	&	0.88640	&	-0.03914		\\
0.8	&	0.88441	&	-0.03009		\\
0.95	&	0.88094	&	-0.02354		\\
0.975 &	0.85384	&	-0.05375		\\
0.9925	&	0.81291	&	-0.04156		\\
 \hline

    \end{tabular}
    \caption{$(l=2)$-led fundamental modes with $M_z=2$ for different scaled background parameter $b/R_H$ for $a/M=0.9756$.}
    \label{tab:a16_n0}
\end{table}

\end{document}